\documentstyle[11pt,aaspptwo]{article}

\lefthead{Le F\`evre et al.}
\righthead{Evolution of the clustering of galaxies}

\begin{document}

\title{The Canada France Redshift Survey VIII:\\Evolution of the 
Clustering of Galaxies from z $\sim$1.}

\author{O. Le F\`evre\altaffilmark{1}}
\affil{DAEC, Observatoire de Paris-Meudon, 92195 Meudon, France\\
lefevre@daec.obspm.fr}

\author{D. Hudon, S.J. Lilly\altaffilmark{1}}
\affil{Department of Astronomy, University of Toronto, Toronto, Canada M5S 1A7}

\author{David Crampton\altaffilmark{1}}
\affil{Dominion Astrophysical Observatory, National Research Council of Canada, Victoria, Canada}

\author{F. Hammer\altaffilmark{1} and L. Tresse\altaffilmark{1}}

\affil{DAEC, Observatoire de Paris-Meudon, 92195 Meudon, France}

\altaffiltext{1} {Visiting Astronomer, Canada-France-Hawaii Telescope, which
is operated by the National Research Council of Canada, 
the Centre de Recherche Scientifique of France and the University of Hawaii}

\begin{abstract}

This paper analyzes the spatial clustering of galaxies in the
Canada-France Redshift Survey (CFRS). 

We have used the projected 
two-point correlation function, $w(r_p)$, to investigate the spatial
distribution of the 591 galaxies with secure redshifts 
between $0 \leq z \leq 1.3$ in the five CFRS 
fields.
The slope of the two-point correlation function for the sample as a whole
is $\gamma=1.64\pm0.05$, very similar to the local slope, and $\gamma$ is therefore not strongly evolving with redshift.
However, the amplitude of the correlation function decreases strongly 
with increasing redshift, so that at $z\approx0.6$ it is a factor of 10 lower
(for $q_0=0.5$)  than for a similarly-selected
local galaxy population, on scales $0.1<r<2h^{-1}$ Mpc ($q_0=0.5$). As a  whole, the CFRS data is adequately represented by $r_0(z=0.53)=1.33 \pm 0.09 h^{-1}$Mpc for $q_0=0.5$, and $r_0(z=0.53)=1.57 \pm 0.09 h^{-1}$Mpc for $q_0=0$.

Unless the galaxy population at high redshift is quite different from any
population seen locally, an unlikely possibility, then this implies
growth of clustering as described by the evolutionary parameter $\epsilon$ to be between $0 < \epsilon < +2$.

No difference in the clustering
of red and blue galaxies is seen at $z \geq 0.5$, although at lower redshifts,
$0.2 \leq z \leq 0.5$,
blue galaxies are somewhat less strongly
correlated than the redder
galaxies, as seen in local samples. This effect could be the equivalent for 
field galaxies 
to the Butcher-Oemler effect
seen in clusters of galaxies. 
The cross-correlation functions between red and blue samples have comparable
amplitudes to the auto-correlation functions of each.

The distribution and power spectrum of pair separations
does not indicate significant periodic patterns in the distribution 
of galaxies along the lines of sight. Finally, 
the densest structures in the survey are identified and characterized,
 including the structure at $z=0.985$ 
in the 1415+52 field, reported previously, and a new
cluster of galaxies at $z=0.78$ in the 1000+25 field.

\end{abstract}


\section{INTRODUCTION}

Deep redshift surveys can provide key information on the evolution of
the spatial distribution of galaxies with cosmic time. In a deep
``pencil-beam'' survey, the line of sight is expected to cross
many structures similar to the sheets, walls and voids identified in
the nearby universe, with  rarer encounters of denser regions
associated with clusters of galaxies. Such surveys can provide information on the existence and properties
of large scale structure in the Universe at high redshift.

It is expected that the clustering of galaxies may well change with epoch
in an expanding universe in which structures evolve and grow
under the action of
gravity. As a consequence, the two-point correlation function $\xi(r)$
should change with cosmic epoch, but the exact form of this evolution
is at present poorly known. 

There have been many studies of the
local $\xi(r)$ in
surveys such as the CfA, Stromlo-APM,  SSRS, IRAS and
others (see e.g. Davis \& Peebles 1983; Loveday et al. 1995, 
Fisher et al. 1994; 
Benoist et al. 1995). They indicate a power-law behaviour with
$\xi(r) = (r/r_0)^{-\gamma}$.
Values for the correlation length $r_0$ range from 3.8 to 7.5
$h^{-1}$ Mpc with a possible dependence on luminosity and galaxy type (Loveday
et al. 1995) and $\gamma \sim 1.7$. 
At higher redshifts, two approaches can be followed to
measure $\xi(r)$. The first is to invert the projected 
angular two-point correlation
function $w(\theta)$ through the Limber equation 
using an observed or predicted redshift distribution N(z) appropriate to the
observed limiting magnitude of the galaxy sample. 
From the angular correlation function
$w(\theta)$, Efstathiou et al. (1991) have shown that faint
galaxies, which are associated with the ``excess'' in the blue number
counts, are rather weakly clustered. At shallower depths, where
spectroscopic surveys are possible,
Hudon and Lilly (1995) inverted a measurement of
$w(\theta)$ using the redshift distribution based on the CFRS and find
a correlation length of $r_0=1.9 \pm 0.1 h^{-1}$ Mpc 
at $z \sim 0.5$.

The alternative approach is to
directly compute $\xi(r)$ from  
the distance information for individual galaxies that is present in
deep redshift surveys.
A direct computation of $\xi$(r) from the redshift 
surveys of Broadhurst et al. (1988) and Colless et al. (1990, 1993) was
carried out by Cole et al. (1994), who did not find any evidence for
evolution in the comoving correlation length for z $<$0.25.
The deep I-band selected
CFRS sample (see Lilly et al. 1995a, CFRS-I; Le F\`evre et al. 1995, CFRS-II;
Lilly et al. 1995b, CFRS-III; Hammer et al. 1995, CFRS-IV; Crampton et al. 1995,
CFRS-V) provides for the first time the opportunity to directly
evaluate the evolution of the two-point correlation function $\xi(r)$
over the redshift range $0 < $ z $< 1$.

Interpretation of apparent changes seen in $\xi(r)$ with epoch
must be approached with caution.
Redshift surveys trace large scale structures by their galaxies. 
Since different 
types of galaxies exhibit different clustering properties at the present
epoch, apparent changes in the clustering of galaxies 
at different epochs may well
reflect a combination of either or both of 
the ``true'' evolution of large scale 
structure and/or changes in the population of galaxies being observed. The
latter can arise either through evolution of the galaxy population or through 
the way the
galaxies are being 
selected at different redshifts. Unfortunately, the evolution of the mix
of different galaxy populations is poorly understood at the present time.  
Thus, any changes observed in the clustering strength
represent the ``apparent'' evolution
in the clustering, and the true growth of structure can only be
determined by making assumptions as to the nature of the galaxy population
dominating the samples at
high redshifts.  A further frustrating ambiguity arises from 
making different assumptions about the value of the deceleration parameter
$q_0$. Assuming a lower value of $q_0$ generally increases the strength
of clustering implied from a given set of observations at high redshift.
This reduces the growth that is implied in clustering up to 
the present epoch, an observational 
reduction that is to a certain degree matched by
the theoretical expectations of reduced growth in low density Universes.

In addition to the small-scale clustering that can be measured with
$\xi(r)$, there has been considerable
interest in the large scale structure information provided by 
deep ``pencil--beam'' surveys since Broadhurst et al. (1990) reported
that the distribution of redshifts in a combination of survey at the
North and South poles up to z$\sim$0.5 is occurring at a preferred scale
of $\sim$128 $h^{-1}$ Mpc.
 
In this paper we first compute $\xi(r)$ from the CFRS sample
by means of the projected 
spatial two-point
correlation function $w(r_p)$ as a function of redshift. We examine
the correlation function for sub-samples of galaxies split by intrinsic 
color, and the cross-correlation function between these.  We then compute and discuss the
auto-correlation function of the galaxy distribution to examine the
galaxy pair distribution vs. comoving distance, and the power spectrum
of the redshift distributions. We finally quantify and examine the
most prominent over-densities seen in the survey.

Except where noted to the contrary, values of H$_0$=100 km s$^{-1}$ Mpc$^{-1}$ 
and q$_0$=0.5 are used throughout 
the paper.

\section{THE CFRS SAMPLE}

The CFRS has been described in detail elsewhere (CFRS I-V). The 
statistically complete subsample consists of
943 objects selected in five 10\arcmin $\times$ 10\arcmin\
fields to have
17.5$\leq I_{AB} \leq$22.5, without regard to color or morphology, and
with minimal surface brightness selection. The sample is 85\% 
spectroscopically identified and this paper is based on the 
591 galaxies in the CFRS that have 
secure redshifts. The redshifts extend up to
z$\sim$1.3, with a median redshift $<z> = 0.56$. 
More than 350 galaxies 
have z$\geq$0.5.  

The I-band selection criterion corresponds to selection in the 
rest-frame 4200 $\leq \lambda \leq 5500 $ \AA ~region for redshifts
0.5 $ \leq z \leq 1$, and thus the
sample of high redshift galaxies
is, in principle, well-matched to  most local samples 
of galaxies. The field dimension of 10\arcmin\ corresponds to 
a comoving dimension of 3.3 $h^{-1}$Mpc at z$\sim$0.5 and to 5.3$h^{-1}$
Mpc at z $\sim$ 1, dimensions comparable
to the $z = 0$ correlation length. 

The CFRS sample thus
provides an unprecedented opportunity
both to investigate the clustering properties of galaxies at different
epochs, and for different intrinsic colors, from a time when the universe 
was $\sim$40\% of its present age, and 
to identify over-densities out to redshifts $\sim$1 (one such
structure was reported
in Le F\`evre et al. 1994).

One significant complication with using the CFRS to study the 
clustering of galaxies is the highly non-uniform spatial distribution of the 
objects selected for spectroscopic study within each field (see CFRS II).
The objects are located primarily in three parallel strips for each of the 5 fields, 
within which almost 100\% spectroscopic sampling was obtained, separated by 
regions where very few spectroscopic observations were carried out.
Our technique for dealing with this uneven spatial sampling
is discussed in \S 3.3. 

The redshift accuracy is 0.0019 r.m.s, or a velocity accuracy of 
550 km s$^{-1}$. To account for this effect in evaluating the evolution of $\xi(r)$ we have used the projected two point correlation function $w(r_p)$ as described in Section 3.2.

\section{THE COSMIC EVOLUTION OF THE TWO-POINT CORRELATION FUNCTION}

\subsection{The epoch--dependent 2--point correlation function $\xi$(r,z)}

The local two-point correlation function is well represented by a 
power--law (Davis and Peebles 1983)

\[\xi(r)=(\frac{r}{r_0})^{\gamma}\]

\noindent
and the epoch--dependent correlation function is usually 
expressed in terms of an evolutionary parameter $\epsilon$ 
(Groth and Peebles 1977; 
Efstathiou et al. 1991),

\[\xi(r,z)=\xi(r,0)\times(1+z)^{-(3+\epsilon)}\]

\noindent Thus, the correlation length can be written

\[r_0(z)=r_0(0)\times(1+z)^{-(3+\epsilon)/\gamma} \]

\noindent To avoid confusion, it should be noted that the correlation 
length $r_0(z)$ is here the correlation length (in physical units) 
that would be measured by a local observer at the epoch in question. 
Thus the 
correlation length will evolve (as $(1+z)^{-1}$) 
even if the clustering pattern 
is fixed in comoving space.

In the case of a clustering pattern fixed in comoving coordinates, 
clustering does not grow with time, and $\epsilon = \gamma - 3 = -1.3$ for $\gamma = 1.7$. When bound gravitational units 
keep a fixed physical size, 
the clustering growth is the result of the increasingly diluted galaxy background 
(it is effectively the voids that are ``growing''), 
and $\epsilon$=0 (Efstathiou 1991; Carlberg 1991). 
For a standard CDM scenario, the mass clustering should grow in the 
linear regime with $\epsilon\sim$0.7  on $\sim$1 h$^{-1}$Mpc scales,  
(Davis et al., 1985). However, clustering 
growth of galaxies in CDM may not follow the linear theory (Brainerd \& Villumsen 1994).

\subsection{Projected 2--point correlation function $w(r_p)$}

To avoid the effect of peculiar velocities and redshift measurement 
errors in evaluating the spatial correlation function $\xi(r)$, 
we have used  the projected function $w(r_p)$ (Davis \& Peebles 1983), 
generalized to the observation of galaxies at high redshifts:
\begin{equation}
w(r_{p})=\frac{c}{H_0\times(1+z)^{2}\times(1+2q_0z)^{1/2}} 
\int_{- \delta z}^{+ \delta z} \xi(r_p,\pi) dz
\end{equation}
with 
\[r_p=(d_{\theta}(i)+d_{\theta}(j))\times tan(\theta_{ij}/2)\]
where $d_{\theta}$ is the angular diameter distance:
\[d_{\theta}=\frac{c}{H_0}\times\frac{q_0z+(q_0-1)(-1+(1+2q_0z)^{0.5})}{q_0^2(1+z)^2} \]
$\theta_{ij}$ is the apparent angular separation between 
galaxy $i$ and galaxy $j$ (degrees). 
Here $\pi$ is the radial distance between the galaxies in physical units. 
The quantity

\begin{equation}
\frac{c}{H_0\times(1+z)^{2}\times(1+2q_0z)^{1/2}} dz
\end{equation}
is the increment in physical distance.  In equation (1), we take the 
$(1+z)^2 (1+2 q_0 z)^{0.5}$ term out of the integral because the effects
of both peculiar motions and redshift errors are constant in velocity space
rather than in physical distance space.  Taking this redshift-dependant
term out means that equation (1) is
valid only when the correlation function is constructed from a sample
of galaxies
occupying only a small
range of redshifts so that the term outside of the integral is roughly 
constant.
With $\xi(r)=(r_0/r)^{\gamma}$ and $r_p<<cz/H_0$, $w(r_p)$ becomes (Davis and Peebles 1983)
\begin{equation}
w(r_p)=\frac{\Gamma(1/2)\Gamma[(\gamma-1)/2]}{\Gamma(\gamma/2)}\times r_0^{\gamma}\times r_p^{1-\gamma}
\end{equation}
($\frac{\Gamma(1/2)\Gamma[(\gamma-1)/2]}{\Gamma(\gamma/2)}=4.0, 4.3 $ ~for $\gamma=1.71$ and $1.64$ respectively)
so that computing  $w(r_p)$ provides a measurement of $r_0$ and $\gamma$.

We have computed $\xi(r_p,\pi)$ with the standard estimator
\[1+\xi(r_p,\pi)=\frac{DD(r_p,\pi)}{DR(r_p,\pi)}\]
where DD is the number of data--data pairs and DR the number of data--random 
pairs (10000 random trials were performed). The projection to $w(r_p)$ 
was then done following equation (1) with $\delta z=0.0075$. Three independently
written codes (by DH, OLF, SJL) have been used for this computation as 
described below.

\subsection{Treatment of non-uniform spatial sampling in the CFRS}

In essence, the correlation function is based on a comparison between the
observed distribution of galaxies and that expected from a random distribution
of points in space, and most estimators of the correlation function are based
on the Monte Carlo generation of large numbers of ``random" data sets.
Unless spatial completeness is achieved, which is extremely costly in 
telescope time at the depth of the CFRS, the use of multi-object 
spectrographs to generate deep redshift surveys 
usually imposes a spatial 
selection function on the final galaxy catalogue relative to the distribution 
of galaxies on the sky.  This selection function arises
due to the various constraints on the positioning of the 
spectroscopic apertures (whether slits or fibres) on the sky.
An obvious example is a possible bias in a single mask
against near neighbours, due to the finite slit length. 
Other more subtle effects are conceivable, including a possible radial 
dependence within the field in the success rate
in securing spectroscopic identifications, etc.
In any event, the spatial selection function must be
properly accounted for in the
generation of the ``random" data sets. 

The observing process of the CFRS restricted the selection of objects to
three parallel strips per field  (CFRS-II). This
geometry was imposed by the mask design which
maximised the multiplexing gain in the spectroscopy by allowing three
non-overlapping spectra in the direction parallel to the dispersion.
When fields were reobserved, emphasis was placed on filling in
objects missed in these strips and on broadening the strips rather than
on observing objects in the areas between the strips. This
resulted in
strips with typical width $\sim1^{\prime}$ and length $\sim10^{\prime}$. 
This strategy
ensured almost 100\% spatial sampling (with virtually no
bias against close pairs of galaxies, CFRS-II) within each strip, but left
large areas with very low ($\sim 0 \%$) spatial sampling. 
Thus, in the final catalogue, while there should be almost no bias
against near neighbours, the distribution of galaxies on arcmin scales 
is highly non-uniform.

This non-uniform spatial distribution of those galaxies identified
spectroscopically 
considerably complicates the generation of the ``random'' data sets
used in the computation of the correlation function. One possible
solution would
have been to generate the random data sets from rectangular areas
matching these strips.  However, in our sample, the details of the
widths and sampling completeness within these strips varied from field
to field due to the differing numbers of masks observed in each field
(see the figures in CFRS-II-IV).  In order to guarantee that the random
data sets had {\it exactly} the same spatial selection function as the
CFRS galaxy sample, each of the random data sets was drawn from {\it
exactly the same ($\alpha$,$\delta$) positions as the overall galaxy
sample} -- i.e. the 591 galaxies with 
secure redshifts (i.e. Confidence Class $\geq$2) in
the statistically complete CFRS sample (see CFRS-V).
In each random data set, these positions were assigned
different, random, redshifts. We adopted three different methods for assigning the random redshifts
and implemented these in three independent codes for computing
$w(r_p)$. These three methods were (a) to use redshifts randomly
drawn from the smoothed N(z) for the whole sample generated from
the luminosity function analysis (see Figure 9 of
 CFRS VI); (b) to use redshifts randomly generated from the observed N(z),
summed over all fields, and binned in intervals of 0.05 in z to smooth
out the picket--fence distribution; and (c)
to use the observed redshifts from the sample with the addition of a
random component in redshift of $\Delta z \leq0.2$.
The results from these three methods  agree to within the nominal
uncertainties obtained from Poisson statistics.
In addition, trials show that adding random offsets in the range 0 to 
10 arcsec to the ($\alpha$,$\delta$) centers for the random data 
sets has no significant effect on the derived $w(r_p)$.

The effect of using the ($\alpha$,$\delta$) positions of the
objects identified spectroscopically for the
``random'' samples is simply to
bias the estimation of the background density of galaxies.
The ``background'' density of galaxies at angular separation $\theta$ is
set by the data itself, and since galaxies are clustered with non-zero
$w(\theta)$, this background density will be biased high, leading
to an underestimate of $w(r_p)$. The effect of this ``random
sampling bias''
is similar to the well known integral constraint
encountered in constructing the projected 2-dimensional
$w(\theta)$ from finite data sets.

The bias in the
background density is simply given by the observed 
amplitude of the projected
2-D correlation function
on the angular scale corresponding to
the value of $r_p$ at the redshift in question.  Thus since

\[ 1 + w = \frac{n_{observed}}{n_{random}} \]

it follows that

\[ 1 + w_{obs}(r_p) = \frac{ 1 + w_{real}(r_p)}{ 1 + w (\theta)} \]

or

\[ w_{real}(r_p) = w_{obs}(r_p)\times(1+w(\theta))  + w(\theta) \]

For $w(r_p) \gg 1 \gg w(\theta)$, as we have, this bias 
is a small angle-dependent multiplicative factor.

In the current analysis, the scale of most interest is $0.1h^{-1} < r
< 2 h^{-1}$Mpc, corresponding to 27$^{\prime\prime}< \theta <$
4\farcm5 at z $\sim$ 0.56, the median redshift of the survey.
Most of the signal is on the larger scales.  The projected
$w(\theta$) on these scales and at these depths is known from large
samples of red-selected galaxies (see e.g., Infante and Pritchet 1995;
Hudon and Lilly 1995). At $19.0 < R < 23.5$, which is broadly
equivalent to our 17.5 $\leq I_{AB} \leq$ 22.5 sample, Hudon and Lilly
(1995) find that $w(\theta)$ varies between 0.01$< w(\theta)
<$ 0.07 on scales 3.3 arcmin down to 27\arcsec. Since our $w(r_p)$ is greater 
than 10 (on all scales at
all redshifts) this bias has a small effect, especially at the larger
scales around 1 $h^{-1}$ Mpc where our S/N is greatest. The bias will
decrease the apparent slope, $\gamma$, by approximately 0.03.  This bias
is sufficiently small that we have chosen to ignore it and have not attempted
to correct for it.

\subsection{Results}

In Figure 1 we show the projected correlation function $w(r_p)$,
computed in the interval $25h^{-1} kpc \leq r_p \leq 2h^{-1} Mpc$  for the
$0 \leq z \leq 1$ CFRS survey. This has a slope $\gamma=1.64\pm0.05$.
This indicates that, while
$\gamma$ is marginally lower than observed in local samples (e.g. $\gamma=1.71$
for the Stromlo-APM redshift survey, Loveday et al., 1995), 
it is not changing significantly
at the mean depth of the CFRS. In fact, the
small difference can be partly accounted for 
by the effect of the random sampling bias 
discussed in \S 3.3. A
similar result has been obtained from $w(\theta)$ reaching similar depth
(Hudon \& Lilly 1995).

In order to examine evolution in the amplitude of the correlation function, the
sample was split into three redshift bins $0.2 \leq z \leq 0.5$,
$0.5 \leq z \leq 0.75$, $0.75 \leq z \leq 1$ (the redshift bins match those of our
analysis of the evolving luminosity function - see CFRS VI). 
The resulting $w(r_p)$ are 
shown in Figure 2.  These have been 
characterized by fitting power-laws with both $r_0$ and $\gamma$ free parameters
and, for uniformity, by fitting a power-law with fixed $\gamma$=1.64, 
following
equation (2).  In the latter case, $r_0$(z=0.34)$=1.83\pm0.18$,
 $r_0$(z=0.62)$=1.10\pm0.15$ and
$r_0$(z=0.86)$=1.05\pm0.10$ are found. 
Errors on $w(r_p)$ over the 
redshift range $0 \leq z \leq 1$ have been derived from the estimates of the correlation function 
obtained in each of the five independent fields of the CFRS, which indicate values
larger than expected from Poisson behavior, compatible
with expectations from simulations (Landy \& Szalay 1993).  
For the computation of $w(r_p)$
in the three redshift bins and in color--redshift bins, error bars on $w(r_p)$ have been 
increased by 50\% from the Poisson value, as the number of 
galaxies in each redshift/color bin
is not sufficient to compute $w(r_p)$ for each individual field. 
Error bars indicated in Figure 1 are from the field-to-field 
variations and are $\approx$50\% higher than the Poisson values.  
In addition, it should be noted that the three independent
codes found essentially identical results, indicating  that
computational errors are negligible.

The above analysis was repeated for q$_0$=0. From equation
(1), the implied $r_0$ is expected to be higher in a lower density universe, both because 
of the $(1+2 q_0 z)^{0.5}$ factor and because the change in angular diameter
distance means a given angular scale corresponds to a larger physical scale.  Indeed,
we find $r_0$(z=0.34)$=2.1\pm0.15$,
 $r_0$(z=0.62)$=1.36\pm0.13$ and
$r_0$(z=0.86)$=1.45\pm0.10$ for the three redshift bins $0.2 \leq z \leq 0.5$,
$0.5 \leq z \leq 0.75$, $0.75 \leq z \leq 1$ respectively.

The sample has also been divided by the intrinsic color of the galaxies,
by dividing the sample into galaxies redder and bluer than the 
Coleman, Wu and Weedman (1980) Sbc spectral energy distribution.
It was shown in CFRS-VI  that the evolution of the
luminosity function of galaxies is quite different for these two color bins. 
In Figure 3,
we show  $w(r_p)$ for the blue and red sub-samples in the redshift
ranges $0.2 \leq $z$ \leq 0.5$ and $0.5 \leq $z$ \leq 0.8$. Fitting again
with a $\gamma = 1.64$ power-law, we find
$r_0$(z=0.34)$= 2.10\pm0.2 $ and $r_0$(z=0.34)$=1.45 \pm 0.25$, and
$r_0$(z=0.65)$=1.21 \pm0.15 $ and $r_0$(z=0.65)$=1.31 \pm 0.15$ 
for the red and blue samples respectively. Finally, within these same
redshift bins, the cross-correlation functions between red and blue
subsamples has also been computed. We find $r_0$(z=0.34, red--blue)$= 1.93\pm0.39 $,  $r_0$(z=0.34, blue--red)$=1.85 \pm 0.41$, 
$r_0$(z=0.65, red--blue)$=0.95 \pm0.10 $ and $r_0$(z=0.65, blue--red)$=1.10 \pm 0.05$. 

Results on the various $w(r_p)$ fits are summarized in Table I. 

\section{DISCUSSION}

\subsection {The evolution of $\xi$(r, z)}

Our data indicates that the correlation length is
decreasing with redshift, with the amplitude of the 
correlation function $\xi(r)$ being lower by a factor $\sim3$ at z$\sim0.6$ compared to z$\sim0.3$.
However, the errors on our $r_0(z)$ points are such that strong constraints
can not be placed on $\epsilon$, the apparent rate of the clustering evolution,
from our data taken by itself.
A formal fit to the 3 CFRS data points shown plotted in Figure 4 yields
$\epsilon=0.4 \pm 1.1$, $r_0(0)=3.3 \pm 1.0$.
Furthermore, as a result of the apparent magnitude 
selection $17.5 \leq I_{AB} \leq 22.5$,
the CFRS data samples different parts of the galaxy
luminosity function at
different epochs (see, e.g., Figure 5 of CFRS VI). In the redshift bins 0.2--0.5, 0.5--0.75, 0.75--1, the
correlation function applies to galaxies with $-17.0 \leq
M(B_{AB}) \leq -20.0$, $-18.0 \leq M(B_{AB}) \leq -21.0$, $-19.0 \leq
M(B_{AB}) \leq -21.0$ respectively (for $h$ = 1). If sub-L* galaxies are less
strongly correlated than L* and super-L* galaxies, 
as reported from local samples (Loveday et al.
1995; Benoist et al. 1995), and if this still holds at higher redshifts,
then, for consistency, our $r_0$ measurement for the $0.2 < z < 0.5$
bin should be revised upwards, by perhaps as much as 50\% (based on the 
trends in Loveday et al. 1995),
leading to a steeper apparent decline with redshift and a change to 
a larger value of the apparent value of $\epsilon$.

In order to determine the best value of $r_0(z)$ representing the whole CFRS sample, we examined all the combinations ($r_0(0), \epsilon$) giving an acceptable fit to the high redshift points. All these ``best fit'' $r_0(0)$ vs. $\epsilon$
curves intersect at $z\sim0.53$ with $r_0(z=0.53)=1.33 \pm0.09 h^{-1}$ Mpc for $q_0=0.5$ and $r_0(z=0.53)=1.57 \pm0.09 h^{-1}$ Mpc for $q_0=0$.

Stronger constraints than formally allowed by the three high redshift $r_0(z)$ points on the apparent value of
$\epsilon$ can be derived when our $r_0$ measurements at high redshift
are taken together with local measurements. For a given value of $r_0(0)$, our data allow to set $\epsilon$ to $\pm0.35$. We can thus see what combinations of ``galaxy population'', parameterized by $r_0(0)$, and
``structure evolution'', parameterized by $\epsilon$, are permitted by the 
CFRS data.
 Davis and Peebles (1983) derived $r_0=5.4 h^{-1}$ Mpc
from the CfA survey limited to m$_B$ $<$14.5, and according to
Fisher et al. (1994), the IRAS sample has
$r_0=3.8 h^{-1}$ Mpc. The correlation length has
been suspected to be dependent on the luminosity and the type of the
galaxies, with  less luminous galaxies being less
strongly clustered than the more luminous galaxies, and early type
galaxies more strongly clustered than late type ones with a range of
correlation length between $2.9 \leq r_0 \leq 7.4 h^{-1}$ Mpc in the 
Stromlo--APM redshift survey (Loveday et al. 1995) and $3.9 \leq r_0 \leq 14.5
h^{-1}$ Mpc for the SSRS2 (Benoist et al., 1995).
The various values
of $r_0$ reported by Loveday et al. (1995) are shown plotted at z$\approx$0
in Figure 4, together with lines representing different values of $\epsilon$.
 
  We discuss here three specific possibilities:

(1) If there was no growth in clustering, i.e., the clustering pattern 
is fixed in comoving coordinates, $\epsilon=-1.3$, then our observed clustering at high redshift
would imply a local value of $r_0(0)=2.0\pm0.2 h^{-1}$Mpc for $q_0=0.5$, $r_0(0)=2.5\pm0.2 h^{-1}$Mpc for $q_0=0$, a clustering strength
much lower than for any local population of galaxies yet observed.
Given the modest changes in the galaxy luminosity function
constructed from this same CFRS sample (CFRS VI and discussion below),
 the existence
of such a currently-unknown  galaxy population is extremely unlikely. 
We therefore conclude that
some growth of clustering {\it must} have 
occurred over this redshift range.

(2) If the physical size of clusters of galaxies had remained fixed
with redshift, i.e., the $\epsilon=0$ case, then 
$r_0(0)=3.0\pm0.2 h^{-1}$ Mpc is implied for $q_0=0.5$ ($r_0(0)=3.9\pm0.2 h^{-1}$ Mpc for $q_0=0.0$). The galaxies we observe at
high redshift in the CFRS would have had to evolve with time into 
the most weakly
clustered (low luminosity) galaxies in, for example, the Stromlo-APM survey (Loveday et al. 1995)
or the SSRS (Benoist et al. 1995). 

(3) Finally, for the CFRS galaxies to be representative of the local 
B-band selected samples with roughly L* luminosities would require
rapid growth of the clustering, with $\epsilon \geq 1$. For instance,
for the CFRS galaxies to evolve into local samples  that have
$r_0 = 5 \pm 0.15 h^{-1}$Mpc (Loveday et al. 1995), would require $\epsilon=2.1 \pm 0.3$ for $q_0=0.5$. 

Given this ambiguity in the interpretation,
what is known about the galaxies that dominate the CFRS sample at 
$0.5 < z < 1.0$ ?  The luminosity function of red 
galaxies in the 
CFRS shows little change over the redshift range of the survey,
while the luminosity function of blue galaxies
indicates
either brightening of the luminosities by $\sim$1.2 mag by z$\sim$0.75,
or a number density evolution of a factor 3 (CFRS VI).  The morphologies 
of the galaxies are now being revealed by deep HST images
(Schade et al. 1995; CFRS-IX). The majority of galaxies appear normal,
representing the same range of Hubble types as seen locally. However, of the
galaxies bluer than Sbc, 30\% show asymmetric structure dominated by 
bright blue regions of star-formation and half have disks that are $\sim$1 mag
brighter than normal galaxies at z = 0.  Thus the population 
at $z \sim 0.6$ is expected to be 
broadly similar to that seen in local B-selected samples of
luminous galaxies, but
with the likelihood of some luminosity evolution in at least part of the 
population.  This suggests that scenarii similar to those
discussed in (2) and (3) above are plausible, probably nearer the latter,
and thus we believe that the ``true'' value of $\epsilon$ 
is likely to be between 0 and +2.

We emphasize that the results presented here  are
made possible by the large number of redshifts, and long baseline in
cosmic epoch
of the CFRS.
Stronger constraints on the evolution of $\xi(r)$ from a direct measurement 
will require  either much larger samples in the
same redshift range, or the direct measurement of $\xi(r)$ at redshifts
significantly in excess of one.   An important gap in the Figure 4 could be 
filled in by analysis of a large sample of luminous galaxies at $z \sim 0.3$. 

\subsection {The effect of $q_0 \sim 0$}

The effect of altering the assumed $q_0$ is straightforward to see in equation (1).
The $w(r_p)$ changes due to both the different incremental comoving 
distance with redshift and the different angular diameter distance.  
Relative to the measurement inferred for $q_0 = 0.5$, 
the effect of decreasing $q_0$ is to 
increase the implied $r_0(z)$ by a factor 

\[
\frac{d_{\theta}(q_0)}{d_{\theta}(q_0=0.5)} \times 
[\frac{(1+z)}{(1+2 q_0 z)} ]^{1/2\gamma} 
\]

For a high redshift sample at $z \sim 0.6$, going from $q_0=0.5$ to $q_0=0$
 has the effect, for fixed
$r_0(0)$, of reducing the implied value of $\epsilon$ by $\sim 1$.

\subsection {The clustering of blue and red subsamples}

Given the known difference between the clustering of blue and red
galaxies in the local Universe (see e.g. Loveday et al. 1995), the
lack of a significant
difference in the auto-correlation functions for blue and red 
subsamples at $z \geq 0.5$ is interesting.  The blue-red and red-blue
cross-correlation functions have comparable amplitudes to the 
auto-correlation functions, suggesting that the blue and red galaxies
are well mixed at $z \sim 0.5$. If, as observed in local samples, early-type galaxies are more strongly correlated than late-type galaxies
(e.g. Loveday et al., 1995), then the fact that we don't see a difference in clustering strength between red and blue samples
for $z\geq0.5$ is 
a strong evolutionary effect. The break-down of the clustering
difference between blue and red galaxies in our CFRS sample
is reminiscent of the
Butcher-Oemler effect in clusters (Butcher and Oemler 1984), 
in which the fraction of blue galaxies in clusters, small at low redshift,
increases with redshift to approach that in the field at $z \sim 0.5$, implying an elimination of the segregation of blue and red galaxies.

Although we have only a small number of galaxies at lower redshifts,
and they are of generally lower luminosities, the difference we see
in the clustering strengths of blue and red galaxies
in our own sample at $0.2 < z < 0.5$ (see Table 1), is consistent with the
picture described above.

The similar auto-correlation and cross-correlation functions of the blue
and red galaxies at $z > 0.5$ argues against a substantial admixture of
any ``new'' population with radically different clustering properties
at this depth, consistent with the indications from the luminosity function
and the morphologies of the galaxies discussed above.

\section{PAIR SEPARATIONS,  POWER SPECTRUM}

The distribution of pair separations (in
comoving coordinates) in the CFRS has been computed for each individual field,
as well as after summing the individual distributions. The
distribution, shown in Figure 5, is quite smooth, with no obvious preferred separation in the combined
data or in the individual fields. This is confirmed through
 power spectrum analysis. The
power spectrum was computed (see e.g., Duari et al. 1992) as
\[P(k)=\frac{1}{N_{gal}} \times |F_N(k)|^2\] with
\[F_N(k)=\sum_{n=1}^{Ngal}exp(-ikz_n)\] There is no statistically
significant peak in the power spectrum in the individual fields or for
the combined data. The galaxy distribution in the CFRS, for a largest
dimension of
$\sim$2500 h$^{-1}$ Mpc, is therefore much smoother than in the
Broadhurst et al. (1990) surveys, and does not exhibit any preferred scale.

\section{SPATIAL DISTRIBUTION OF GALAXIES} 

Localized peaks in the redshift distribution can be due either to
dynamically bound clusters of galaxies with typical scales of
the order of one Mpc, or dense filaments with scales $\approx$10 Mpc. The
CFRS survey has the ability to characterize the distribution of
galaxies on transverse scales of $4 h^{-1}$ Mpc at the median redshift of the
survey. Since the redshift histograms of the five CFRS fields are highly
suggestive that the lines of sight are crossing a large number of
galaxy over-densities, this section aims to identify the most
significant over-dense regions observed in our survey, and evaluate the type
of spatial distribution of galaxies in these
structures.

\subsection{Density enhancements in N(z) distributions}

The galaxies are distributed in  redshift with a ``picket--fence''
distribution as identified in shallower surveys. We show in Figure 6
that this type of distribution extends over the full dimension of the
survey or $\sim$2500 h$^{-1}$ Mpc, with prominent peaks obvious out
to z$\sim$1.

In order to identify significant density enhancements in the redshift
distributions, we used the following approach. First, the local
overdensity of galaxies was computed at redshift intervals of
0.001 summing over galaxies within $\pm1500$km s$^{-1}$, corresponding
to the typical spread in velocity of clusters of galaxies:\\
\[Galaxy~ over-density = \] \\
\[ \sum_{-1500}^{1500}(N_{gal}-N_{gal,mean}) /
\sum_{-1500}^{1500} N_{gal,mean} \]\\ where $N_{gal,mean}$ is the mean
N(z) distribution derived from our observed luminosity function (CFRS-VI), normalized to the number of galaxies with redshifts
in each field. Then, we have defined a ``signal/noise'' for each peak
as  \[S/N = \frac{N_{gal,3000}-N_{gal,mean}}{\sqrt{N_{gal,3000}}}
\] All peaks with an over-density larger than 2.5 and $S/N\geq$2.5
have been retained and examined in detail. The seven peaks detected in this
manner are listed in Table 2. 
 %(We have repeated this analysis for a
%velocity interval of 500 km/s and find...).
 Figure 6 presents the observed N(z) distribution, the ``smoothed'' N(z) 
predicted from the CFRS luminosity function, and flags the 7 significant 
peaks detected above. Figure 7 shows the projected distribution of the 
galaxies in these peaks for each field. 

\subsection{Projected Galaxy Density Enhancements}

To check for possible correlations between the over-densities in
redshift and over-densities in the projected spatial distribution of
galaxies, the projected galaxy density has been computed in each of the
fields in the following way. On each point of a grid of 500$\times$500 pixels,
each pixel being 1\farcs24  square, the projected density has been
computed for galaxies with 17.5$\leq I_{AB} \leq$22.5 
following the prescription by Dressler (1980)
\[D_{gal,proj}=10/\pi \times r_{10}^{2}\]  
The projected density maps are
presented in Figure 8, after normalisation to the mean projected
density in each field, and overlaid on the galaxies identified from the
photometry (CFRS-I). No projected density enhancements
 with\\
$D_{gal,proj}/D_{mean,proj} \geq 3.5$\\
 are seen in our magnitude range.

%-> Compare to Postman cluster search.

\subsection{Individual Density Enhancements}

\subsubsection{Structure at z=0.985 in 1415+52}

The most significant density enhancement in the redshift distribution
is in the 1415+52 field, with an over-density of 12 and a S/N of 3.2 at
a redshift z=0.985. This structure has been described in Le F\`evre et
al. (1994), and is the strongest single density enhancement
detected in all of the CFRS survey. One of the most remarkable properties
for such a high over-density is the relatively uniform distribution of
galaxies in this field as is evident in Figure 7. Further studies of
this structure over a wider field are continuing.

\subsubsection{A cluster at z=0.78 in 1000+25}

The second largest over-density in the redshift distribution
is found in the 1000+25 field with an
over-density of 5 and a S/N=3. The galaxies identified in this
structure are peaked around a number of close galaxies around galaxy
CFRS10.1991 at $\alpha_{2000}$=10$^h$00$^m$31$.^s$64,\\
$\delta_{2000}$=+25$^{\circ}$17\arcmin11\arcsec, as shown in Figure 8. The
projected density of galaxies with magnitudes between that of
the brightest galaxy and $I_{AB}$=23.5
shows a well-defined maximum  with a peak density excess of
3.5, as shown in Figure 9. Within 0.5 h$_{50}^{-1}$ Mpc, there are 37
galaxies brighter than $I_{AB}$=22.6 ($\sim$ m$_3+$2), while only
$\sim$13 are expected based on the average CFRS counts. The N$_{0.5}$ parameter
defined by Bahcall (1981) is therefore N$_{0.5}$=24, indicative of a
moderately rich cluster. The observed velocity dispersion of 910 km 
s$^{-1}$ is in good agreement with that predicted from the relationship
 between $N_{0.5}$ and the radial velocity dispersion
(Bahcall, 1981), 960km s$^{-1}$. Of the 12 galaxies with redshifts in
this cluster (+1 in the supplementary catalog), 10 have $V-I$ colors
redder than the color of a  redshifted Sbc spectral energy distribution 
according to Coleman et al. (1980). This
is also consistent with these galaxies being located in a region 
with a density comparable to that of rich clusters. An analysis of the
spectrophotometric properties of galaxies in this cluster will be
presented elsewhere.

This cluster would have been detected from the projected galaxy density
for 17.5$\leq I_{AB} \leq$23.5 only if the over-density threshold had been set
lower than  3.5$\times$mean background. This demonstrates that searching for
clusters at very high redshifts from the projected galaxy density alone
is very challenging and, in general,  only very rich clusters will be selected.

\subsubsection{Large scale over-dense region in 0000-00}

Examination of the redshift distribution for the 0000-00 field shows a
broad peak at z$\sim$0.25. The distribution of galaxies in this peak is
not identified by the above algorithm to be a significant over-density
within a box $\pm$1500 km s$^{-1}$. However, the detection algorithm
identifies an over-density of a factor 5.3 with S/N=3.1 when the
exploration box is $\pm$3500km s$^{-1}$. If this is confirmed by additional
redshift measurements in this sparsely sampled field, this could be the
indication that this pencil beam is crossing an over-density of galaxies 
with a typical scale of  70 h$^{-1}$ Mpc.

\subsubsection{Other over-dense peaks}

As shown in Figure 6, the galaxies in the other peaks of the redshift distributions
do not have any preferred spatial locations, and are more
or less evenly distributed over the fields. For example, the third highest
peak, in the 0300+00 field, has about 16 galaxies producing an
over-density of 4.4 in the redshift distribution, but no apparent
concentration in the spatial distribution. This probably 
indicates that, given the typical cross-section of our survey fields
at the redshift of the peaks, $\sim$5 h$^{-1}$ Mpc, our pencil beams
are crossing sheets of galaxies similar to what is seen locally.

\subsubsection{Galaxy distribution as a function of V-I color in over-dense regions}

Figure 10 shows the redshift distribution of galaxies redder or bluer
than a  present day Sbc galaxy in all our fields.
 Although the numbers are small, it is
interesting to note that only the z=0.78 structure in the 1000+25 field
has a predominantly red galaxy population (10 ``red'' vs. 2 ``blue''
galaxies), while the other structures show a roughly equivalent number
of ``red'' and ``blue'' galaxies. This is consistent with the observed
segregation of galaxy types in which early-type galaxies favor the
regions of highest density  while late-type galaxies are found
preferentially in low density regions (Dressler, 1980). The peaks 
observed in our survey, apart
from the z=0.78 peak in the 1000+25 field identified with the cluster
described above,
sample regions which have lower densities than the rich cluster environments
for which this effect is observed.

\subsubsection{QSOs}

Of the 6 QSOs identified in the CFRS sample, none is identified in a
peak of projected galaxy density (see fig.2). Of the 3 QSOs at
redshifts $\leq$1.2, only the QSO 14.1303 is associated with a
significant overdensity in the redshift distribution (Le F\`evre et
al. 1994). A more detailed description of the QSO environments is
given in Schade et al. (1995, CFRS-X).

\section{SUMMARY}

We have established the following results from the analysis of the
spatial clustering of galaxies in the CFRS survey:\\ 
(1) The slope of
the projected correlation function $w(r_p)$ computed for the whole
CFRS sample, i.e., at a mean redshift z$\approx$0.6, is $\gamma=1.64 \pm 0.05$,
very similar to the local slope, and $\gamma$
is therefore not strongly evolving with redshift. \\ 
(2) There is clear evidence for evolution of the clustering amplitude 
with redshift. At z$\approx$0.6, the amplitude of $\xi(r)$ is $\sim$10 times
lower than locally on scales $0.1< r < 2 h^{-1}$Mpc ($q_0=0.5$).\\
(3) The
correlation length at $z=0.53$ is $r_0=1.33 \pm 0.09h^{-1}$Mpc for $q_0 = 0.5$.
This is inconsistent with any known 
local population of galaxies if the clustering is fixed in comoving space;
it is consistent with only the most weakly clustered local galaxies in the
$\epsilon = 0$ weak growth case, and requires strong growth, 
$\epsilon \geq +1$ to match local B-selected samples of luminous galaxies.
Our analyses of the evolving luminosity function (CFRS-VI) and of HST images
of CFRS galaxies (CFRS IX) indicate that the population mix cannot have
changed dramatically, hence growth of the clustering properties in the range
$0 \leq \epsilon \leq 2$ must have occurred.\\
(4) The computation of the correlation length with $q_0=0$ decreases the apparent value of the $\epsilon$ evolutionary parameter by $\sim 1$. At $z=0.53$ the CFRS sample as a whole is best described by $r_0=1.57 \pm 0.09h^{-1}$Mpc for $q_0 = 0$.\\
(5) There is no difference in the
clustering of blue and red galaxies at redshifts above 0.5, while there
seems to be a difference in clustering amplitude of 2.5 in the range $0.2
\leq z \leq 0.5$, similar to what is observed in local samples. 
This may be analagous to the Butcher-Oemler effect 
in clusters at similar redshifts.\\ 
(6) The distribution of galaxy
pair separations in each of the fields, and in the full sample, does not
exhibit any regular pattern similar to the 128 $h^{-1}$ Mpc pattern found by
Broadhurst et al. (1991).\\
(7) The two strongest peaks in the redshift distributions are identified 
with a large wall of galaxies at z=0.985 (Le F\`evre et al. 1994) and with 
a newly-identified  moderately rich cluster of galaxies at z=0.78.\\

{\bf Acknowledgements}

We are grateful to C. Benoist, A. Blanchard, F. Bouchet, R. Carlberg, S. Maurogordato,
and P.J.E. Peebles for useful discussions in the course of this work. We thank the referee, J. Loveday, for his review of the paper.
The research of JDH and SJL are supported by NSERC of Canada, and
we acknowledge travel support from NATO. 

\clearpage

\begin{table*}
\begin{center}
\caption{Results on $w(r_p)$}
\begin{tabular}{llll}
 Redshift & N$_{gal}$ &  $r_0$ & $\gamma$ \\ 
\tableline
0 -- 1 & 565 & 1.5$\pm$0.15 & 1.64$\pm$0.05\\
0.2 -- 0.5 & 186 &  1.83$\pm$0.18 & 1.64 (fixed) \\
0.2 -- 0.5 & 186 &  2.10$\pm$0.15 & 1.58$\pm$0.09 \\
0.5 -- 0.75 & 196 & 1.1$\pm$0.15 & 1.64 (fixed) \\
0.5 -- 0.75 & 196 & 1.36$\pm$0.13 & 1.67$\pm$0.16 \\
0.75 -- 1 & 130  & 1.05$\pm$0.1 & 1.64 (fixed) \\
0.75 -- 1 & 130  & 1.45$\pm$0.1 & 1.69$\pm$0.12 \\
0.2 -- 0.5 (red) & 93  & 2.1$\pm$0.2 & 1.64 (fixed) \\
0.2 -- 0.5 (blue) & 93 & 1.45$\pm$0.25 & 1.64 (fixed) \\
0.2 -- 0.5 (red--blue) & & 1.93$\pm$0.39 & 1.64 (fixed) \\
0.2 -- 0.5 (blue--red) & & 1.85$\pm$0.41 & 1.64 (fixed) \\
0.5 -- 0.8 (red) & 96  & 1.21$\pm$0.15  & 1.64 (fixed)\\
0.5 -- 0.8 (blue) & 139 & 1.3$\pm$0.15 & 1.64 (fixed) \\
0.5 -- 0.8 (red--blue) & & 0.95$\pm$0.09  & 1.64 (fixed)\\
0.5 -- 0.8 (blue--red) & & 1.10$\pm$0.05 & 1.64 (fixed) \\
\end{tabular}
\end{center}
\end{table*}

\begin{table*}
\begin{center}
\caption{Details of the highest peaks in the redshift distribution}

\begin{tabular}{lclccr}
 Field & z$_{peak}$ & N$_{gal}^a$ & Over-density$^b$ & S/N & $\sigma_{v}$  \\ 
\tableline
0300+00 & 0.217 & 10 (12) & 2.5 & 2.5 & 510  \\
0300+00 & 0.612 & 16 (17) & 4.4 & 3.4 & 1060  \\
0300+00 & 0.707 & 13 & 3.4 & 2.9 & 900  \\
1000+25 & 0.466 & 12 (14) & 3.6 & 2.9 & 700  \\
1000+25 & 0.778 & 12 (13) & 5.0 & 3.0 & 910  \\
1415+52 & 0.746 & 11 (12) & 3.2 & 2.7 & 812  \\
1415+52 & 0.986 & 12 (15) & 12.0 & 3.2 & 873  \\
\end{tabular}
\end{center}
\tablecomments{$^a$ Numbers in brackets include galaxies from the supplementary catalog, 
$^b$ Over-densities as computed for $\pm$1500 km $s^{-1}$}
\end{table*}

\newpage

\begin{figure}

\caption{The projected two-point correlation function $w(r_p)$ for 
the galaxies in the CFRS sample with $0\leq z \leq 1$. Error bars are a direct 1--$\sigma$ estimate from the 
field-to-field variations. The slope of the two point correlation function at the mean redshift of the survey $z\sim0.56$ is $\gamma=0.64\pm0.05$, and therefore $\gamma$ is not strongly evolving compared to local galaxy samples. The amplitude of $w(r_p)$ is $\sim$10 times lower at z$\sim0.56$ than for local galaxy samples  with $r_0\sim5 h^{-1}$Mpc.}

\caption{The projected two-point correlation function $w(r_p)$ for 
three redshift ranges 0.2--0.5, 0.5--0.75, 0.75--1. Measurement errors 
have been increased by 50\% from the Poisson errors. There is clear evidence for an evolution of the correlation strength with redshift within our sample, as well as compared to local samples with $r_0\sim5 h^{-1}$Mpc. In the three increasing redshift ranges, the CFRS galaxies have absolute luminosities $-17.0 \leq
M(B_{AB}) \leq -20.0$, $-18.0 \leq M(B_{AB}) \leq -21.0$, $-19.0 \leq
M(B_{AB}) \leq -21.0$ respectively (for $h$ = 1). If the correlation length varies with galaxies luminosities (Loveday et al., 1995), our $r_0$ measurement for $0.2\leq z \leq$0.5 should be revised upward, possibly by up to 50\%, 
for consistency.}

\caption{The projected two-point correlation function $w(r_p)$ for the galaxies
 redder than an unevolved Sbc galaxy (filled squares) and bluer than Sbc 
(open squares). The two populations have indistinguishable properties in the high redshift bin, which indicates a strong evolution when compared to the strong
difference in the clustering strength between early and late-type galaxies observed locally (Loveday et al., 1995).} 

\end{figure}

\newpage

\begin{figure}

\caption{The evolution of the correlation length with redshift. 
Filled squares represent values of $r_0$ computed for $q_0=0.5$, 
 empty squares are for $q_0=0$, dotted vertical lines indicate the 
range of redshifts of each $r_0$ value. The short dashed line is
 for $\epsilon=0$, the dotted line for $\epsilon=-1.3$, the long dashed line for $\epsilon=0.7$, and the dot-dash line for $\epsilon=2.1$ (see text). Values spanning the range of  local correlation lengths reported for various samples in Loveday et al. (1995) have been indicated. The derivation of $r_0$ from  $w(\theta)$ by Hudon and Lilly (1995) has been indicated by circles (filled: $q_0$=0.5, open: $q_0$=0). Coupled to the local measurements, and our knowledge of the properties of galaxies at high redshift, our data indicates that moderate to strong growth of the clustering is required ($0 \leq \epsilon \leq 2$).}

\caption{Distribution of pair separations in comoving Mpc from all CFRS
fields combined. There is no apparent periodic pattern.} 

\caption{Redshift distribution in the five CFRS fields, in redshift bins $\delta z=0.007$. The mean galaxy density derived 
from the CFRS luminosity function (CFRS-VI) is indicated as a full line. 
The 7 peaks with a galaxy over-density $\geq$2.5 and a S/N$\geq$2.5 (see text) are 
identified by diamonds atop the peaks.}

\caption{Projected distribution of galaxies in each of the peaks identified by the 
detection algorithm. Apart from the galaxies in the cluster at z=0.78 in 1000+25, most 
galaxies in the peaks are distributed evenly within the measurement area (cf. fig.6).}

\end{figure}

\newpage

\begin{figure}

\caption{Projected 2D galaxy over-density maps superimposed on the galaxies identified 
in photometry in the range 17.5$\leq I_{AB} \leq$22.5 (ellipses). Blackened ellipses indicate the galaxies with a redshift identified from spectroscopy. The projected galaxy 
density has been computed with $D_{gal,proj}=10/\pi \times r_{10}^{2}$, and then 
normalized to the mean projected galaxy density in each field. The lower right panel shows the distribution of normalized projected galaxy density in each field. The largest projected over-density is $\sim3.5$ times the mean galaxy background. } 

\caption{Cluster of galaxies at z=0.78 in the 1000+25 field. The field shown is 
120x120 arcsec$^2$ around $\alpha_{2000}$=10$^h$00$^m$31.64$^s$,
 $\delta_{2000}$=+25\deg 17\arcmin 11\arcsec. The projected galaxy over-density is 
shown as contours starting at $1.5\times$ the mean density, with spacings of 0.4, 
and it peaks at an over-density of 3.5 on the two brightest galaxies in the concentration. Thirteen galaxies have confirmed redshifts within several $km s^{-1}$ of the brightest cluster galaxy.} 

\caption{Redshift distribution of galaxies  redder than a redshifted present day Sbc  (shaded) compared to the redshift distribution of all galaxies (open). The mix of blue and red galaxies is relatively even in the high and low density regions, except for the z=0.78 peak in 1000+25, consistent with it being associated to a cluster of galaxies.}
\end{figure}

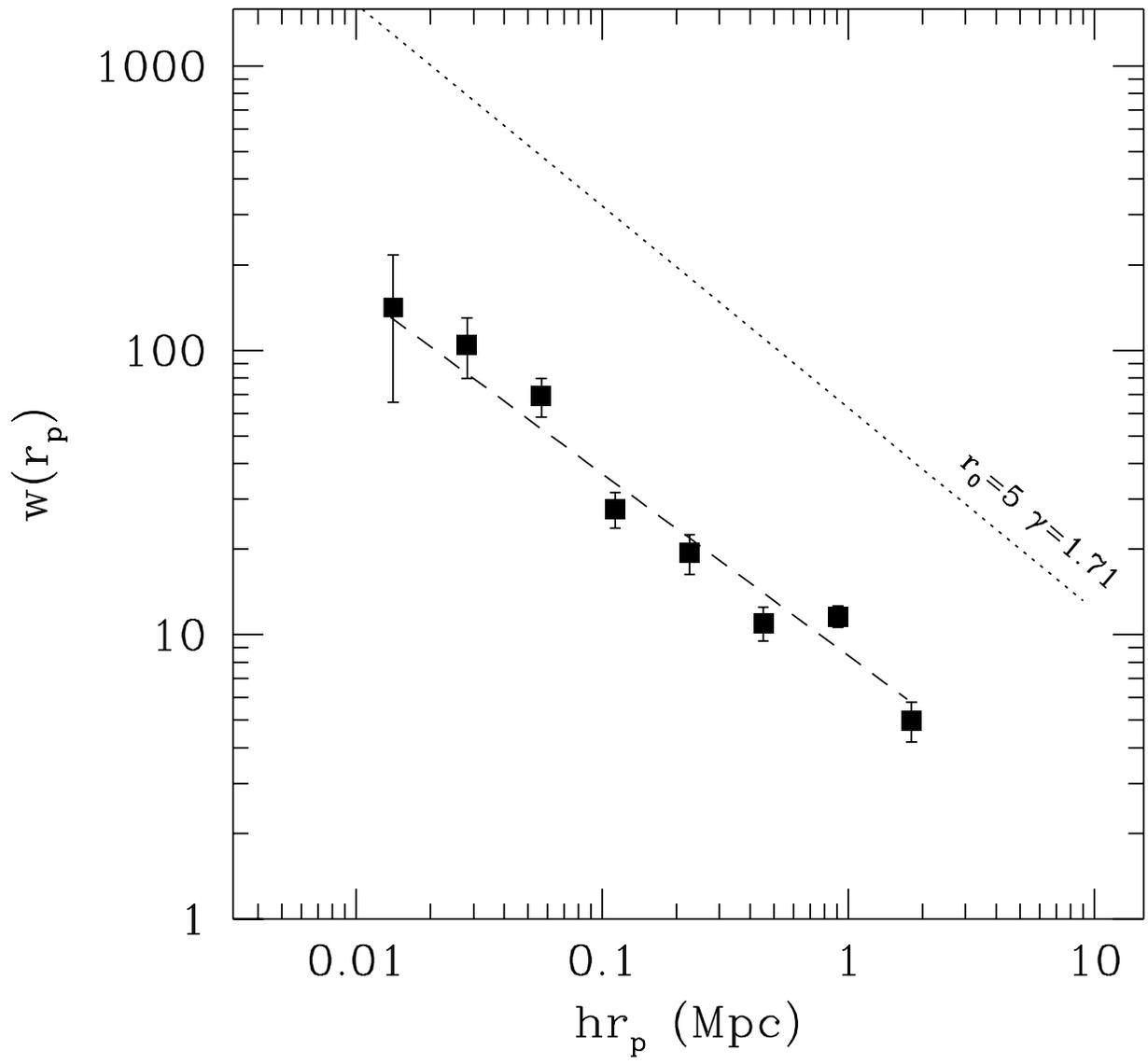
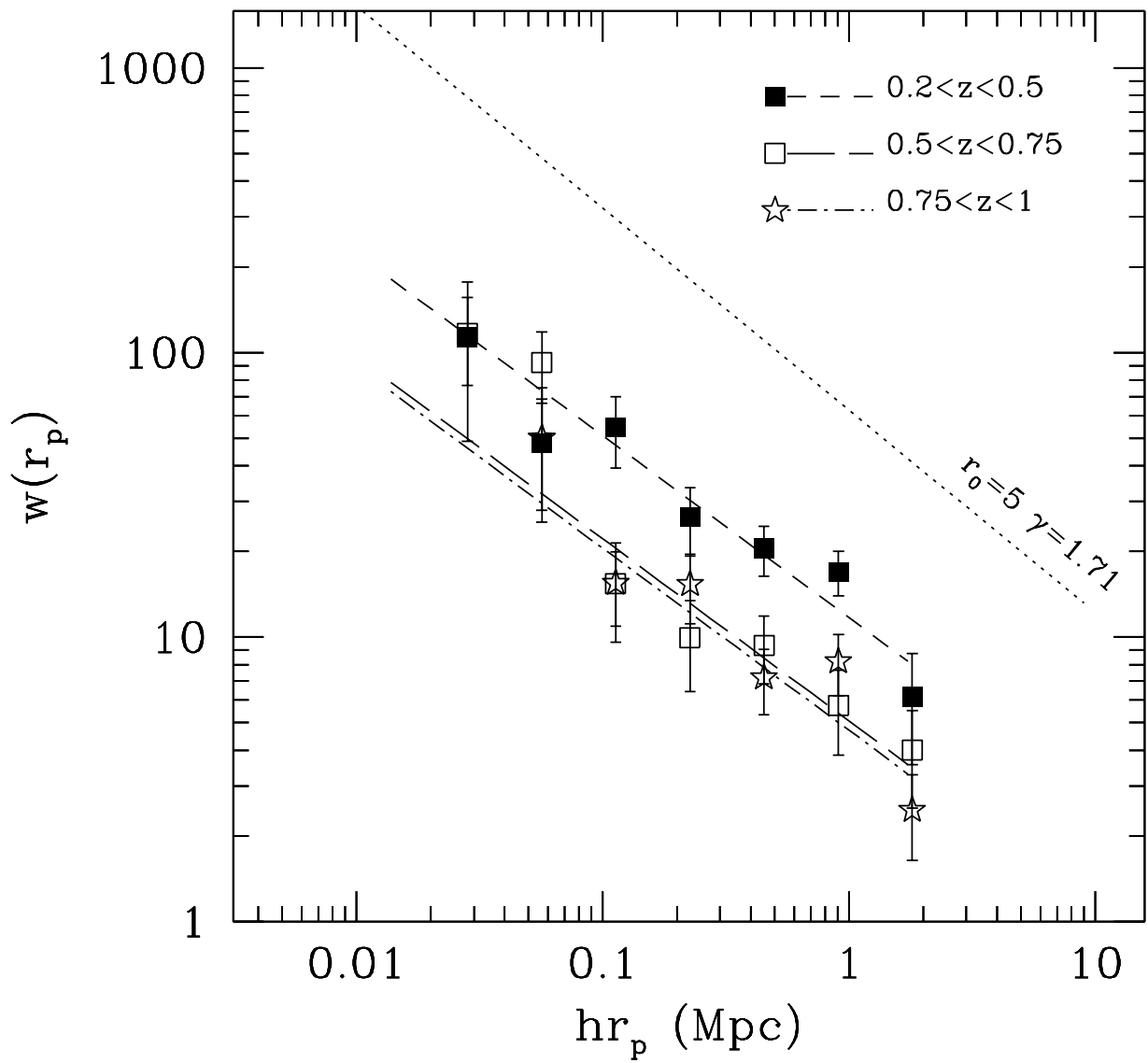
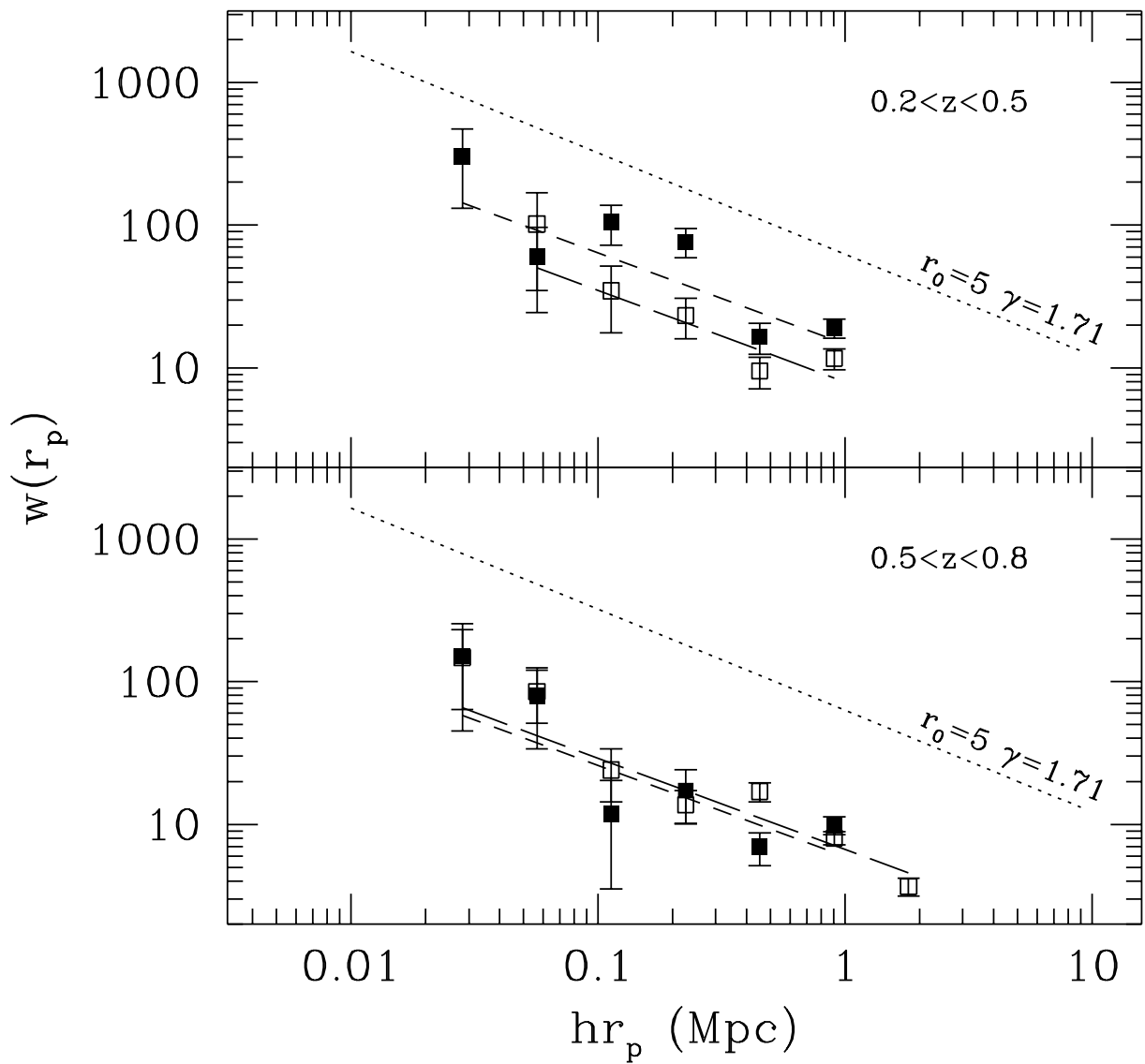
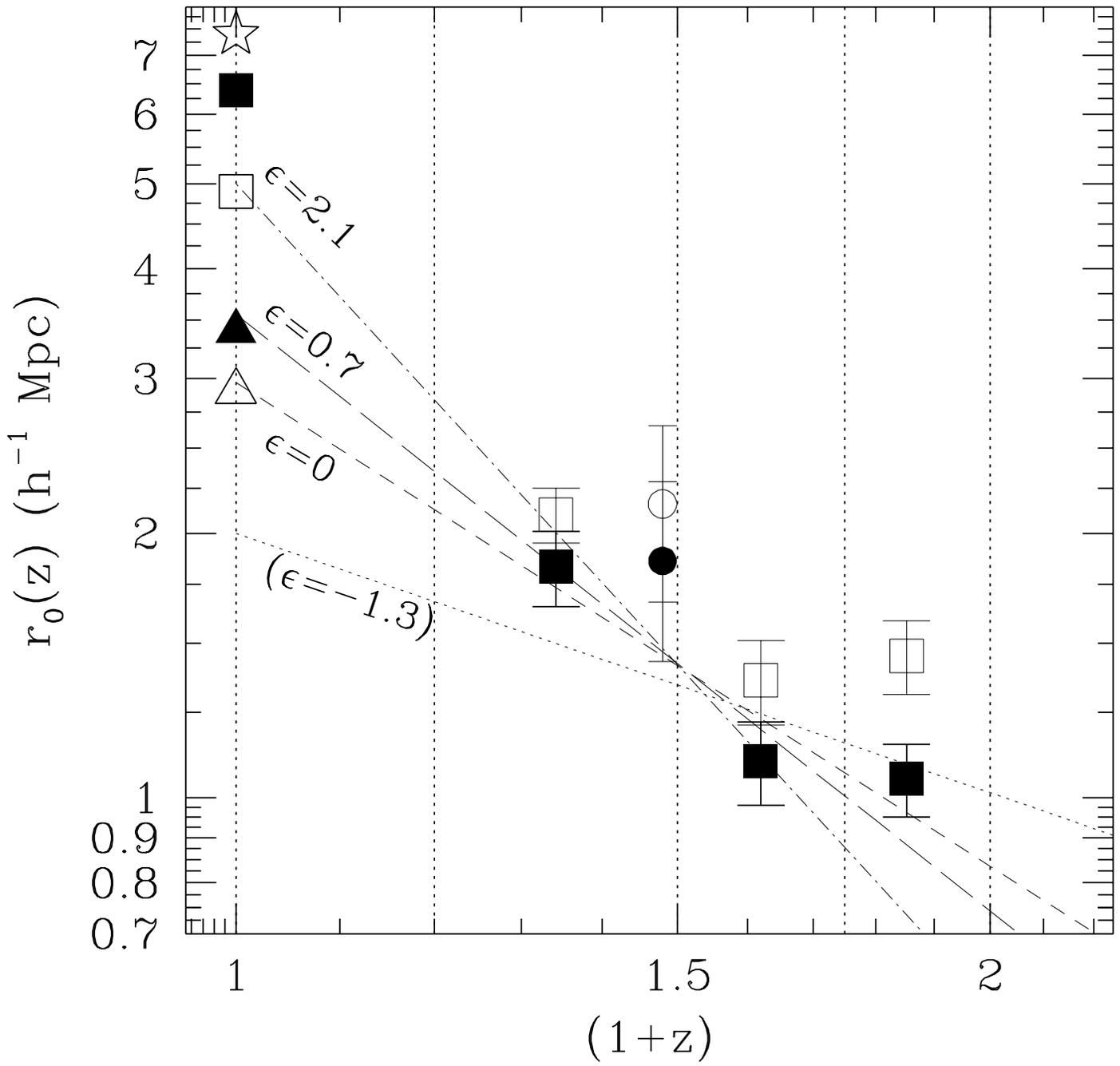
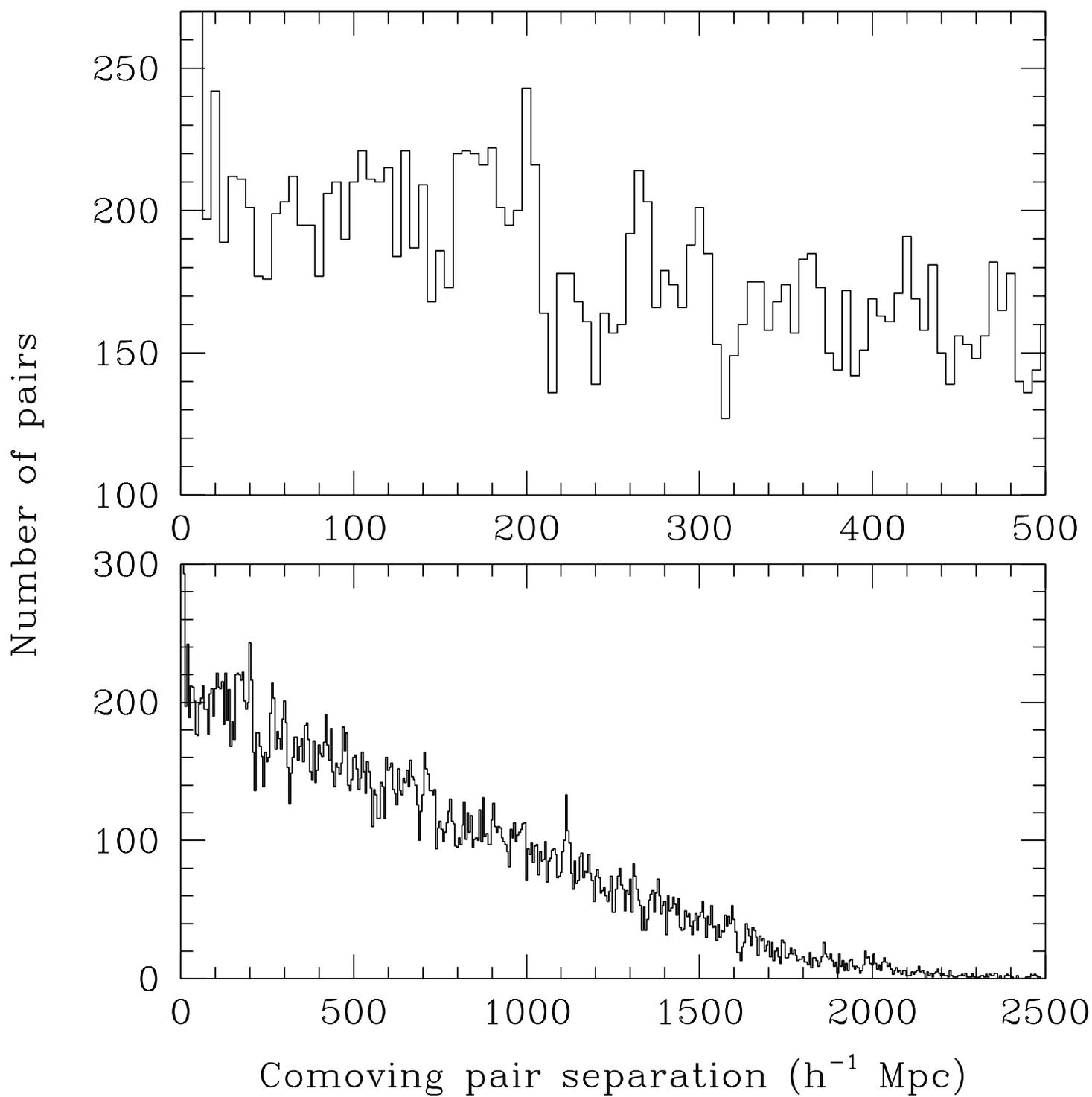
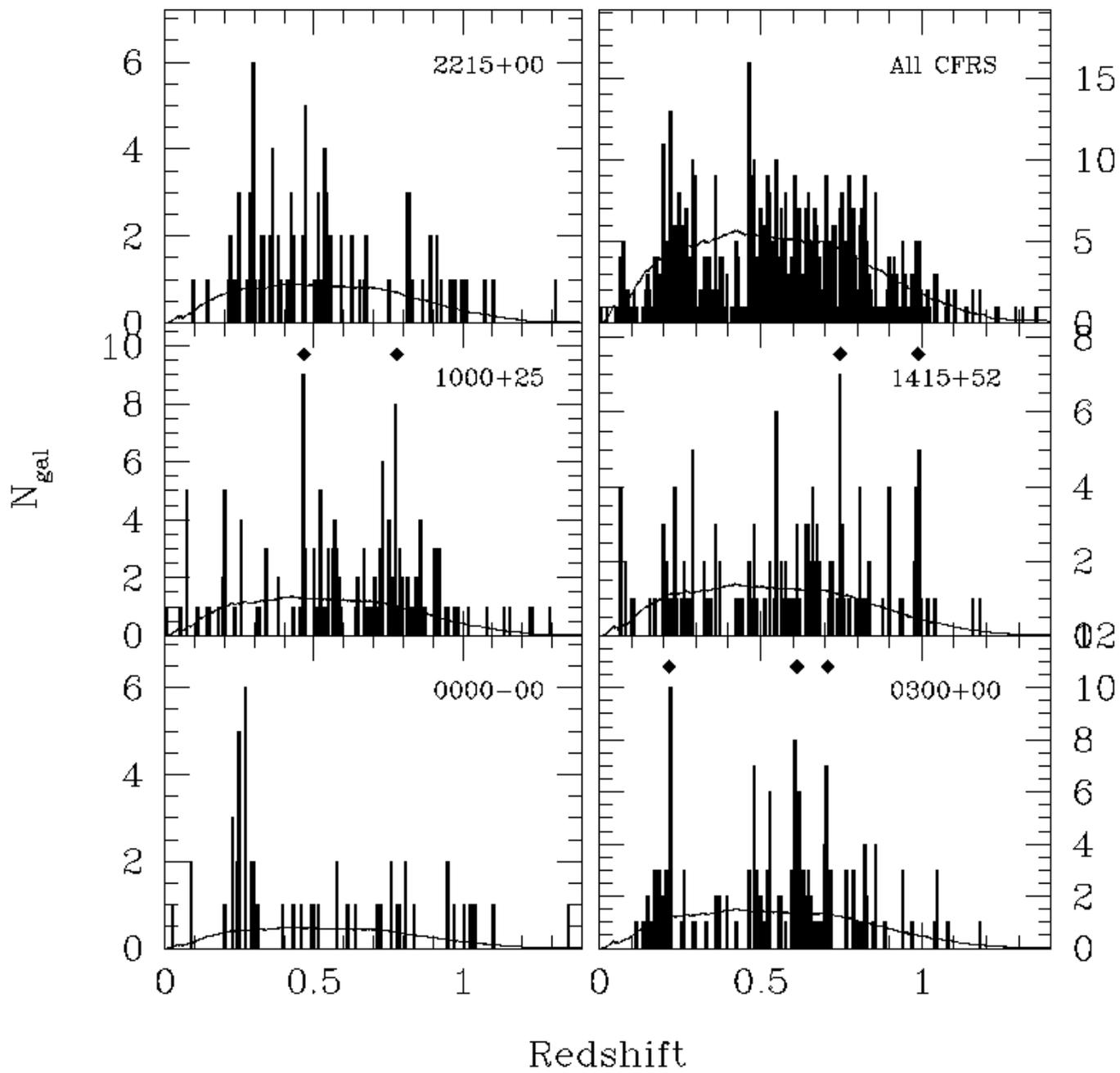
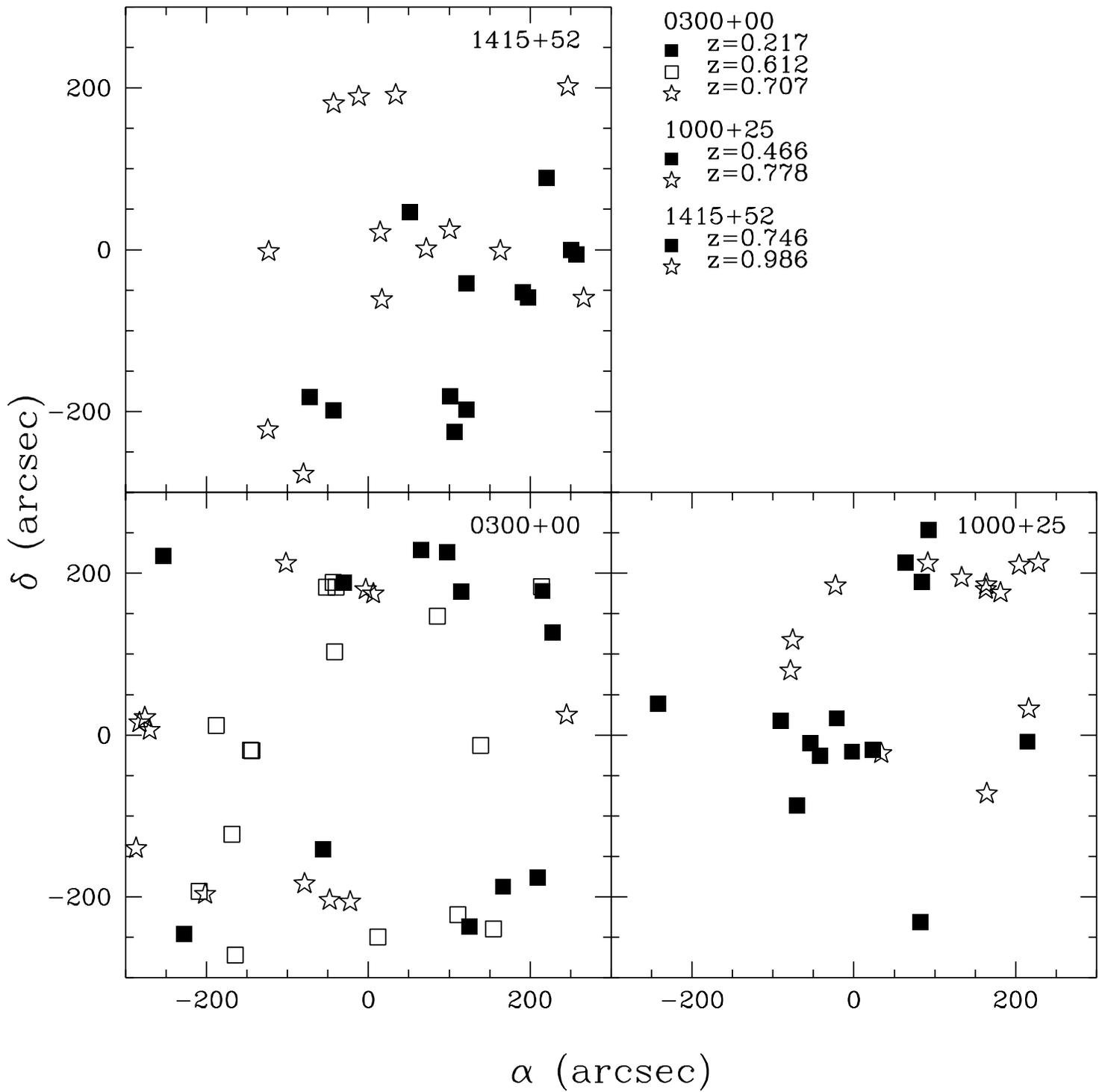
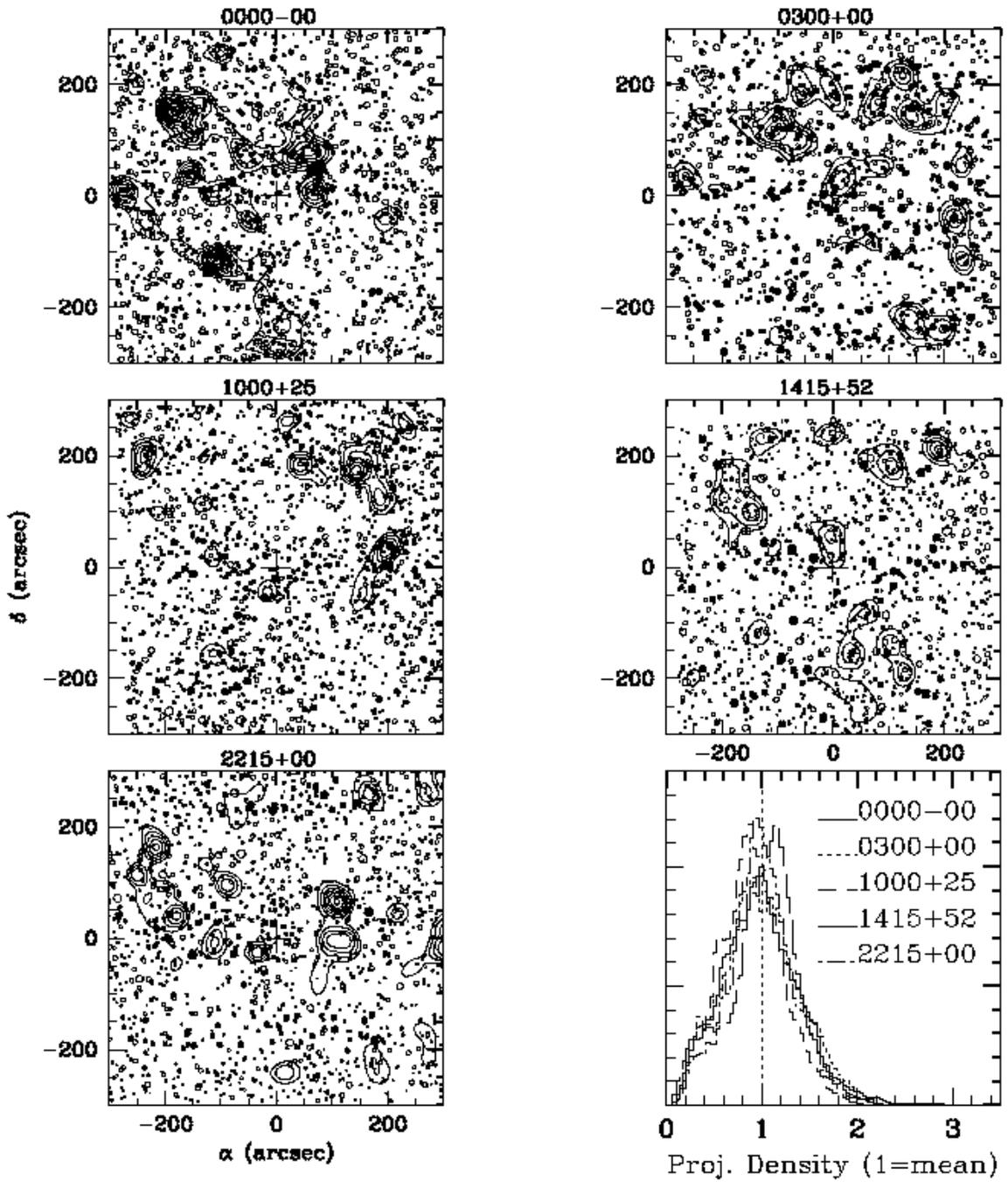
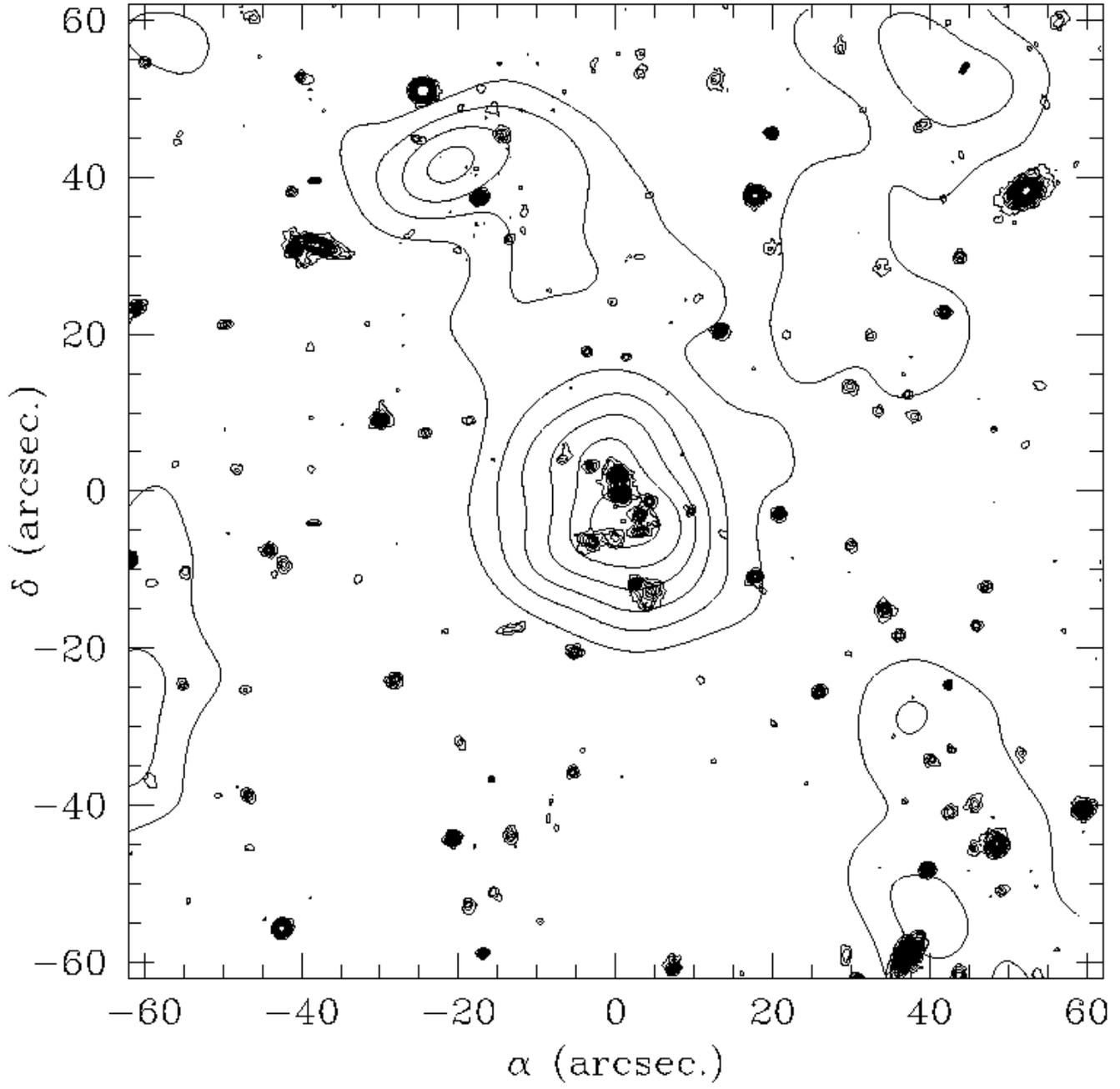
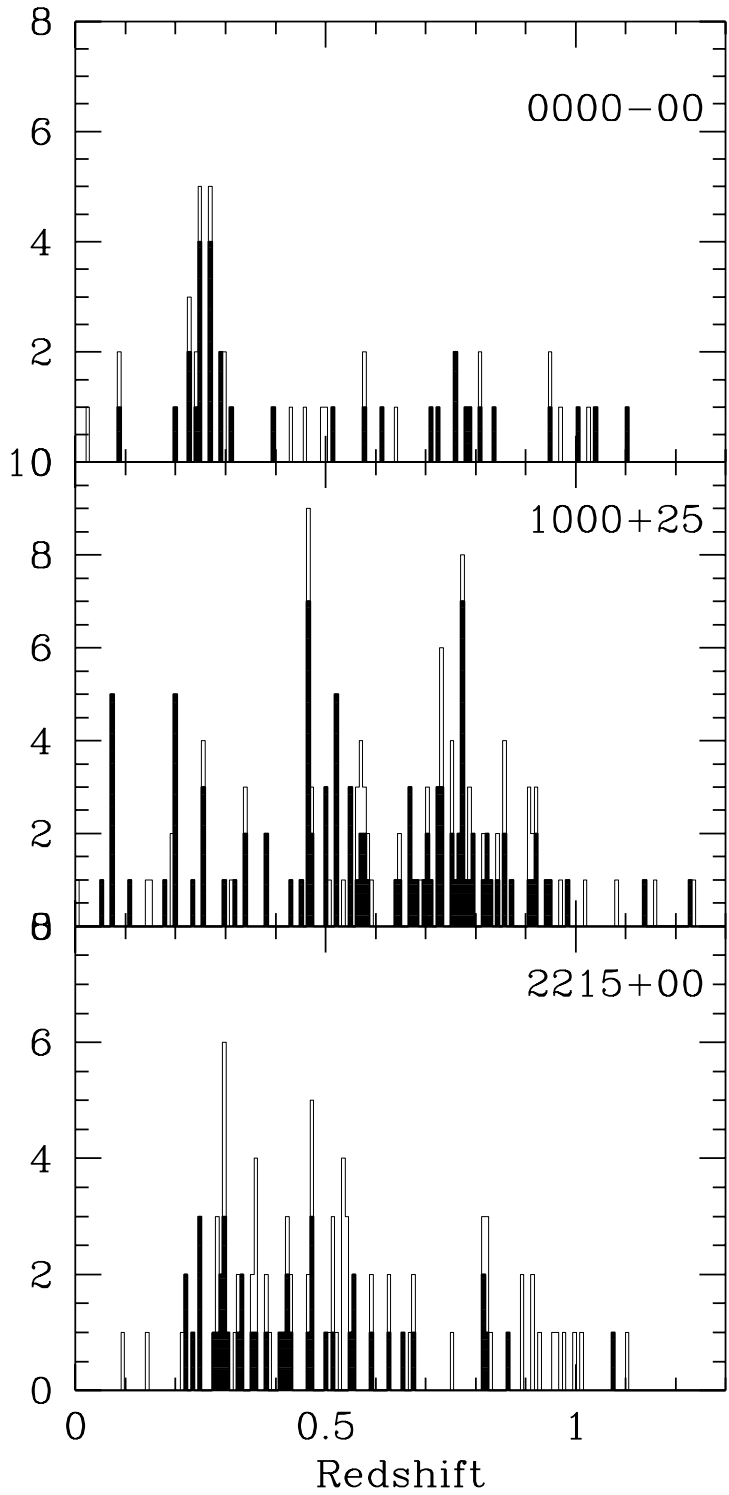
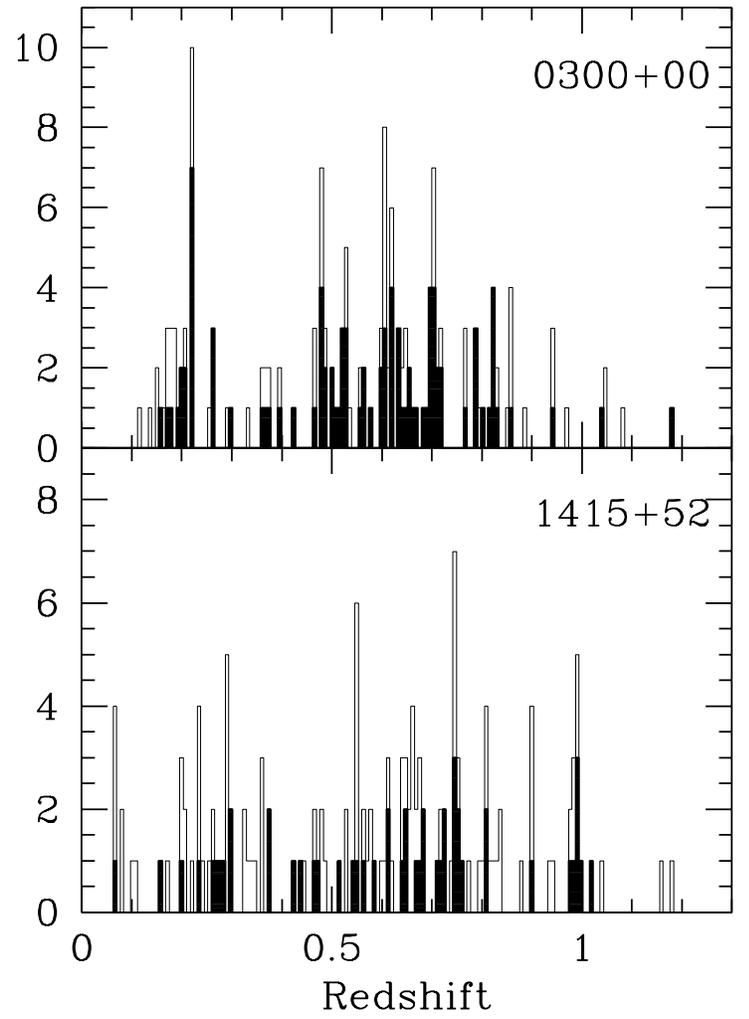
\end{document}